\documentclass[lettersize,journal]{IEEEtran}
\usepackage{amsmath,amsfonts}
\usepackage{algorithmic}
\usepackage{algorithm}
\usepackage{array}
\usepackage[caption=false,font=normalsize,labelfont=sf,textfont=sf]{subfig}
\usepackage{textcomp}
\usepackage{stfloats}
\usepackage{url}
\usepackage{verbatim}
\usepackage{graphicx}
\usepackage{cite}
\begin{document}

\title{Chess variation entropy and \\engine relevance for humans}

\author{Marc Barthelemy\\
Universit\'e Paris-Saclay, CNRS, CEA, Institut de Physique Th\'eorique, 91191, Gif-sur-Yvette, France\\
Centre d’Analyse et de Math\'ematique Sociales (CNRS/EHESS) 54 Avenue de Raspail, 75006 Paris, France}




\maketitle

\begin{abstract}
  Modern chess engines significantly outperform human players and are
  essential for evaluating positions and move quality. These engines
  assign a numerical evaluation $E$ to positions, indicating an advantage
  for either white or black, but similar evaluations can mask varying
  levels of move complexity. While some move sequences are
  straightforward, others demand near-perfect play, limiting the
  practical value of these evaluations for most players. To quantify
  this problem, we use entropy to measure the complexity of the
  principal variation (the sequence of best moves). Variations with
  forced moves have low entropy, while those with multiple viable
  alternatives have high entropy. Our results show that, except for
  experts, most human players struggle with high-entropy variations,
  especially when $|E|<100$ centipawns, which accounts for about $2/3$
  of positions. This underscores the need for AI-generated evaluations
  to convey the complexity of underlying move sequences, as they often
  exceed typical human cognitive capabilities, reducing their
  practical utility.
\end{abstract}


\section{Introduction}

The study of chess offers a fascinating intersection between computational science and complex systems analysis. With its finite rules but near-infinite possibilities, chess serves as an ideal environment for developing and testing algorithms. The game’s deterministic nature allows for the clear application of computational techniques, from heuristic searches to neural networks, providing valuable insights into problem-solving and optimization. Chess has long been foundational to computer science research, marked by milestones such as IBM's Deep Blue defeating Garry Kasparov in 1997 \cite{Campbell}. Modern AI models like AlphaZero \cite{Sadler:2019, Grath} further highlight chess’s significance in advancing computational methods \cite{Silver:2018}. Beyond its mathematical structure, chess also serves as a psychological battleground where players must anticipate opponents’ moves while concealing their own strategies, adding layers of complexity beyond the board. In addition, the recent rise of online chess platforms has enabled large-scale data analysis, introducing concepts from statistical physics and complex systems \cite{Perotti:2013, DeMarzo:2022, Blasius:2009, Maslov:2009,Ribeiro,Schai:2014, Schai:2016,Atash,Chow:2023,Barthelemy2023}. Researchers have identified patterns such as power-law distributions in opening move frequencies, reflecting the self-similar nature of the game tree \cite{Blasius:2009, Maslov:2009}. Long-range memory effects in game sequences, which vary according to player skill levels, have also been observed \cite{Schai:2014, Schai:2016} and the impact of chess experts' knowledge was discussed in \cite{Chassy:2011}. Additionally, the response time distribution in rapid chess was analyzed in \cite{Sigman}, and the decision-making process of chess players was examined in \cite{Chacoma}.

Mathematically, a chess game can be represented as a decision tree where each branch ultimately leads to a win, loss, or draw \cite{Shannon:1950}. This model is particularly useful in the middlegame, where the objective is to find a move that leads to a favorable branch with a majority of winning paths. Conversely, the endgame often demands precise calculation, as deviating from the optimal path can result in a draw or loss. The evolution of computer chess, from the early algorithms by Shannon and Turing (as documented in \cite{Levy:1988}) to modern AI, highlights chess’s enduring significance in computational research. This is especially true given the vast number of possible games that make exhaustive search impractical \cite{Tromp:2022, Steinerberger:2015}. The resurgence of interest in chess, driven by AI engines that surpass human abilities, has revitalized the study of openings and strategic concepts. These engines not only renew chess theory but also serve as a prime example of human-computer interaction, demonstrating how AI can enhance human understanding. As such, chess provides an excellent testbed for developing concepts in explainable AI \cite{Gunning:2019}.
\begin{figure}[h!]
	\includegraphics[width=0.4\textwidth]{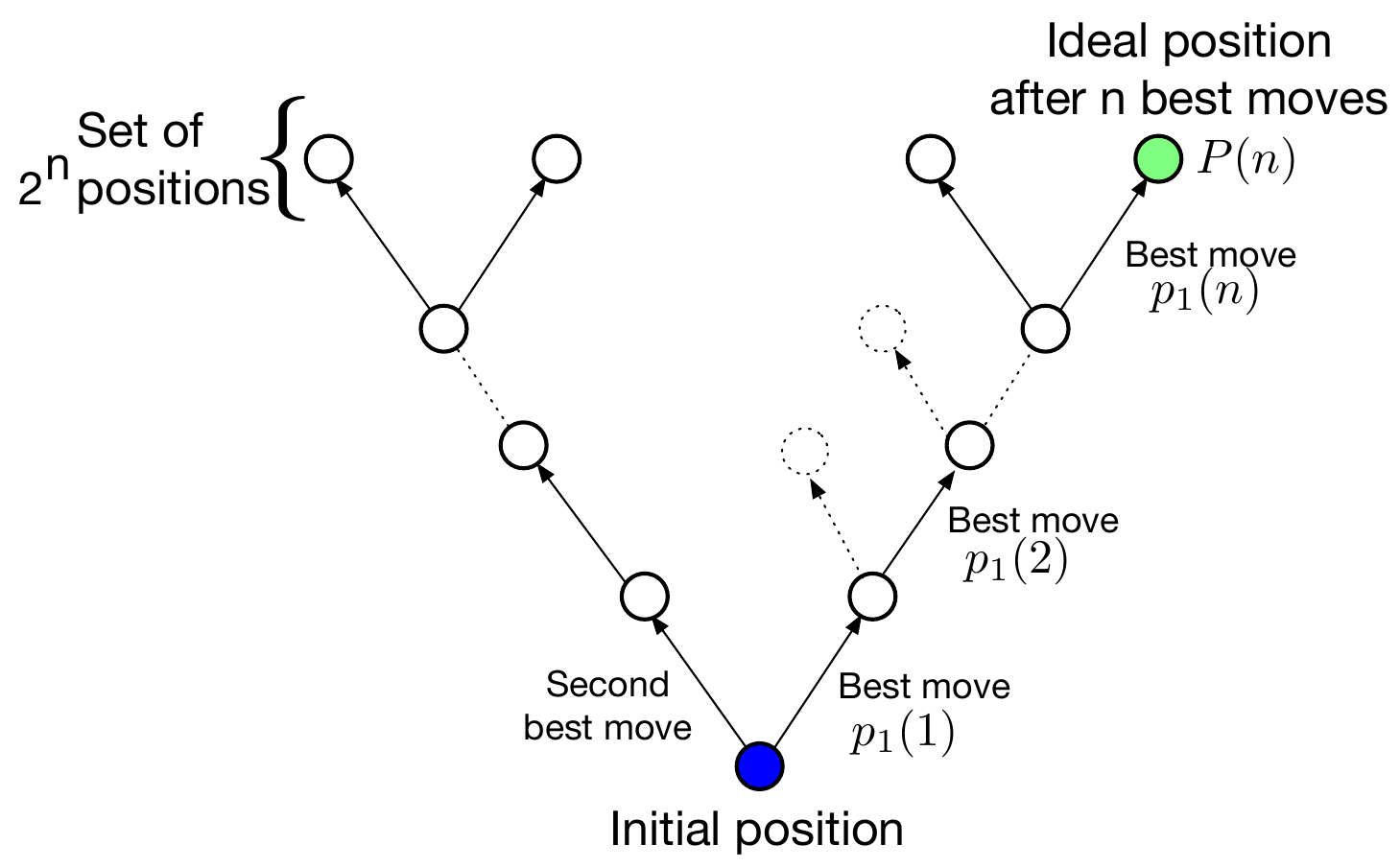}
	\caption{From a given position, the path towards the best (winning) position necessitates to find the best moves consecutively. The probability to reach the ideal position after $n$ half-moves is $P(n)$ (Eq.~\ref{eq:Pn}).}
	\label{fig:path}
\end{figure}

The challenge, then, is how to navigate this decision tree and identify the best sequence of moves, particularly in the middlegame, where calculations become nearly impossible due to the explosive combinatorial complexity. The need for a scientific approach to chess has long been advocated by many players, notably Richard R\'eti \cite{Reti}, a prominent proponent of hypermodernism, who proposed a more systematic approach to the game. Today, widely accepted principles—such as control of the center, control of key squares, colors, material imbalance, and pawn structure—represent the initial steps toward a scientific theory of chess. Modern engines assist in evaluating the quality of moves by analyzing a given position and providing a numerical evaluation $E$, measured in centipawns (the unit of advantage in chess, where 100 centipawns is equivalent to the value of one pawn). This evaluation is positive if White is favored and negative if Black is favored. This information is crucial as it allows players to test various moves and observe the resulting changes in $E$. These evaluations are linked to a sequence of optimal moves known as the principal variation. However, the practical utility of these evaluations can be limited, particularly when the sequence of moves is difficult for non-expert players to discern. This paper examines the relevance of these evaluations for human players and assesses their practical applicability.

\section{Entropy of a variation}

We analyze a sequence of moves (also called a variation) from a given position in the following way. We will consider a depth of $n$ half-moves (or plies) which corresponds to $n/2$ moves.  A given position $P$ has an evaluation $E(P)$, and the best move maximizes the evaluation of the new position $P'$ (for white, while for black $E(P')$ is minimized). We denote by $E_1$ the evaluation for the best move, by $E_2$ the evaluation for the second best move (for white we have $E_1>E_2$ and for black $E_1<E_2$), and $E_i$ the evaluation for the $i^{th}$ best move. We define the following `gap' \cite{Barthelemy2023}
\begin{align}
  \Delta= E_1-E_2
\end{align}
This quantity $\Delta$ characterizes the difference of quality between the best move and the second best move, and has been shown to display a wide range of values \cite{Barthelemy2023,Chacoma}. When $\Delta$ is large, the best move is significantly better than the second best move and many players won't miss it. When $\Delta$ is small (typically less than $100$ centipawns), it could be more difficult to find the best move. We therefore characterize a player's strength by their ability to detect the best move, quantified by the parameter $\Delta_0$. Specifically, if $\Delta > \Delta_0$ the player can identify the best move; however, if $\Delta < \Delta_0$, the player's resolution is insufficient to determine the best move between the top two options. It is natural to assume that the probability to make the best move of the usual logit form in discrete choice theory (see for example \cite{Train} and references therein)
\begin{align}
  p_1=\frac{\mathrm{e}^{E_1/\Delta_0}}{\sum_{i=1}\mathrm{e}^{E_i/\Delta_0}}
\end{align}
We will consider the first two moves only and the probability to find the best move is then
\begin{align}
  p_1=\frac{1}{1+\mathrm{e}^{-\Delta/\Delta_0}}
\end{align}
The sequence of $n$ half-moves is then characterized by the probability $P(n)$ to reach a winning position given by
\begin{align}
  P(n)=\prod_{i=1}^np_{1}(i)
  \label{eq:Pn}
\end{align}
where $p_{i}(i)$ is the probability to find the best $i^{th}$ half-move (see Fig.~\ref{fig:path} for an illustration of this point). 
The corresponding information entropy  \cite{Shannon1948} is
\begin{align}
  S=-\log_2\prod_{i=1}^n
  \frac{1}{1 +\mathrm{e}^{-|\Delta(i)|/\Delta_0}}
  \label{eq:entropy}
\end{align}
where $\Delta (i)=E_{bm}(i)-E_{2bm}(i)$ is the gap corresponding to the position at half-move $i$ ($E_{bm}(i)$ is the best move at ply $i$, and $E_{2bm}(i)$ the second best move). For a sequence of $n$ plies, the entropy is then $0\leq S\leq n$. This quantity $S$ is the information quantity
that a player has to process when visualizing the next $n$ half-moves. Note that the player has to find his own
best moves but also the ones of his opponent. If $|\Delta(i)|\gg \Delta_0$ for all half-moves, then $S=0$: the choice is clear at each step. In contrast, if $|\Delta(i)|\ll \Delta_0$ -- which corresponds to the worst case --  we have the largest possible entropy $S=n$. We note that this quantity should not be confused with the entropy of move distributions as discussed in \cite{Chassy:2011}: in this work the authors weighted each move by a factor reflecting how often it appeared in their database. Here in contrast, we discuss the entropy of a sequence of moves. 

The concept behind Eq.~\ref{eq:entropy} is that when evaluating a variation, the moves that must be ruled out represent the information that needs to be processed and temporarily stored in working memory \cite{Baddeley}. As a result, the entropy measure given by Eq.~\ref{eq:entropy} can be interpreted as a proxy for the cumulative cognitive load associated with a variation, akin to the overall load experienced during a task \cite{Paas}. For instance, when considering entropy as a measure of uncertainty in binary decisions, an entropy value of $4$ bits corresponds to $2^4 = 16$ equally probable outcomes. For humans, this represents a moderate level of uncertainty, implying $16$ equally likely possibilities—a number that is within the cognitive capacity of the human mind. In the context of a binary decision tree, this can be visualized as selecting the correct path through $4$ half-moves, leading to one of 16 different terminal positions. An entropy of $4$ bits is thus within the manageable range for human cognition. In fact, research \cite{Miller} has shown that humans generally find entropy levels around $2-3$ bits ($4-8$ equally likely choices) to be most comfortable to process. Higher entropy values, however, can become overwhelming, depending on the task. Here, we will consider a scenario involving $8$ half-moves, equivalent to $4$ full moves, and an entropy of $4$ bits implies that half of these plies are nearly forced, leaving at most $16$ possible outcomes to consider. As entropy increases (more choices), the cognitive load also rises, making decision-making more challenging.

\begin{figure}
  \includegraphics[width=0.5\textwidth]{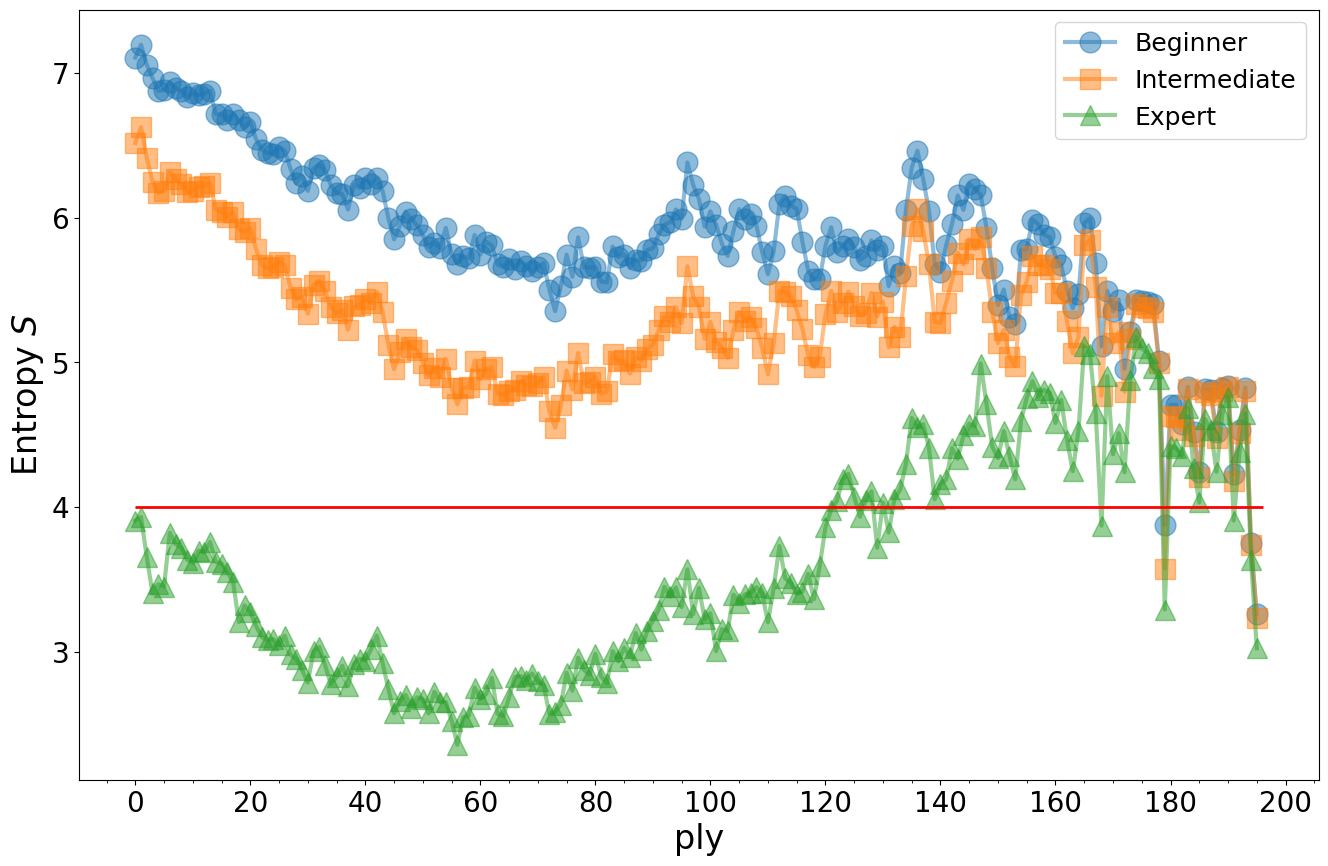}
  \includegraphics[width=0.5\textwidth]{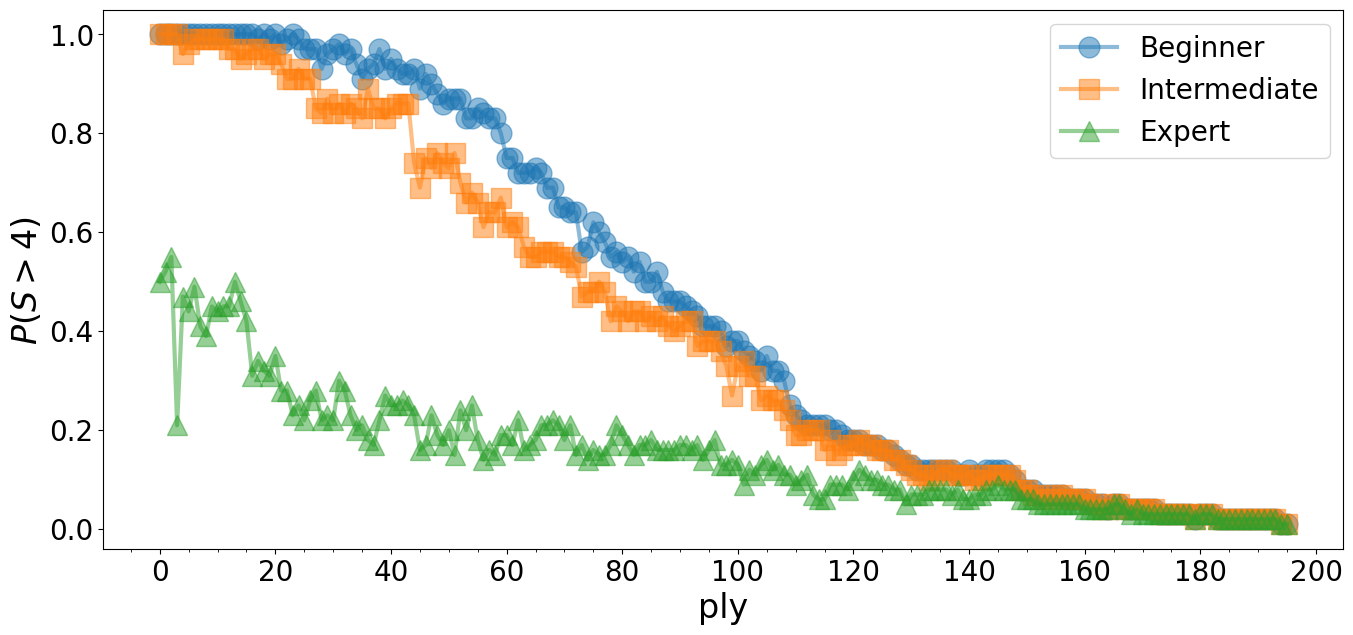}
        \caption{(Top) Evolution of the average entropy $S$ versus ply (or half-move) for different levels of players (beginner, intermediate, expert).
          (Bottom) Evolution of the probability to observe an entropy larger than $4$ during the game, computed for different levels. 
          These results are obtained for an average of 100 games played at the World Rapid Chess Tournament 2023 (Samarkand).}
	\label{fig:Svsply}
\end{figure}

We computed this entropy averaged over $100$ games taken at the World Rapid Chess Tournament 2023 (Samarkand). For $\Delta_0$, we
can estimate from \cite{Chacoma} (using the fact that for $\Delta=\Delta_0$, $p_1\approx 0.73$) that the order of magnitude is $\Delta_0\sim 10^1$ for experts and $\Delta_0\sim 10^2$ for beginners. Based on this, we selected $\Delta_0=100$ for beginners, $\Delta_0=50$ for intermediate players, and $\Delta_0=10$ for experts. Typically, this implies that beginners struggle to differentiate between moves even when one wins a pawn or more, while experts can detect subtle differences in move quality. We also verified that varying these values yields similar results. The evaluation was provided by the Stockfish engine \cite{SF} (see Methods). The average entropy for three different player skill levels during the game is shown in Fig.~\ref{fig:Svsply} (the ply represents the progression of the game is plotted on the x-axis).

We observe that the opening phase is challenging for all players, but experts quickly adapt, reducing the complexity of the positions they face. In general, the entropy for all player levels decreases as the game progresses from the opening phase (low ply $\leq 20-30$) to the middlegame (mid-range ply). However, there is a noticeable increase in entropy towards the endgame (high ply $\sim 100$), particularly for intermediate and expert players. The experts’ entropy is notably low between ply $20$ and $100$, indicating they are more comfortable with the middlegame, where the skills of experts are the most relevant. Their ability to detect the best move significantly reduces the entropy of positions faster than beginners and intermediates. The rise in entropy beyond ply $100$ suggests increasing complexity or critical points in the endgame where precision is crucial. We also observe that there is a clear separation between the entropy levels for beginners, intermediates, and experts throughout the game, with beginners consistently experiencing the highest entropy and experts the lowest. For beginners, the entropy is consistently high throughout the game (with $6<S<7$). This suggests that beginners face a high level of complexity or uncertainty when selecting moves, regardless of the game stage. For intermediates, the entropy starts slightly lower than beginners, and decreases more steadily, reaching a minimum before rising again towards the end of the game. This indicates that intermediate players have a moderate level of uncertainty, which fluctuates more during the different phases of the game. Experts exhibit the lowest entropy, starting around $4$ and decreasing further in the early game before rising again towards the endgame. This shows that experts can generally handle complexity better, but still face increasing difficulty as the game progresses, particularly in later stages. The fact that expert players’ entropy is often below $4$, particularly in the early to mid-game, suggests that they can maintain control over the game with manageable complexity. In contrast, beginners and intermediates consistently operate above this threshold, indicating that they face higher cognitive loads.

In Fig.~\ref{fig:Svsply}(Top) we represented the average entropy, but there are variations from a game to another one. In order to extract information from these fluctuations, we show in Fig.~\ref{fig:Svsply}(bottom), the probability $P(S>4)$ to observe a value of entropy larger than 4. These results are consistent with those discussed above for average values. In particular, we see that for experts, there is a rapid decline of high-entropy situations, underscoring their superior ability to navigate the game's complexities, particularly in the critical middlegame phase.  The drop in $P(S > 4)$ for all levels indicate that complex situations requiring a large cognitive load are more seldom. There are therefore significant differences in how chess players of varying skill levels manage the complexity of positions throughout a game. Experts excel at reducing complexity as the game progresses, particularly in the middlegame and endgame, while beginners and intermediates continue to face challenging, high-entropy positions for much longer. The persistence of high entropy for non-experts underscores the importance of tailored training and simplified recommendations to help them navigate through complex positions.

\section{Engine evaluation and human players}

For a given board position $P$, engines provide an evaluation $E(P)$. By analyzing the entropy of the principal variation, we can assess how relevant this evaluation is for human players. To do this, we calculated the engine evaluation $E(P)$ and the corresponding principal variation entropy $S_{pv}$ for each position $P$ across 100 games from the World Rapid Chess Championship 2023 in Samarkand, Uzbekistan (which corresponds to about $9.500$ positions).

\begin{figure}
  \includegraphics[width=0.5\textwidth]{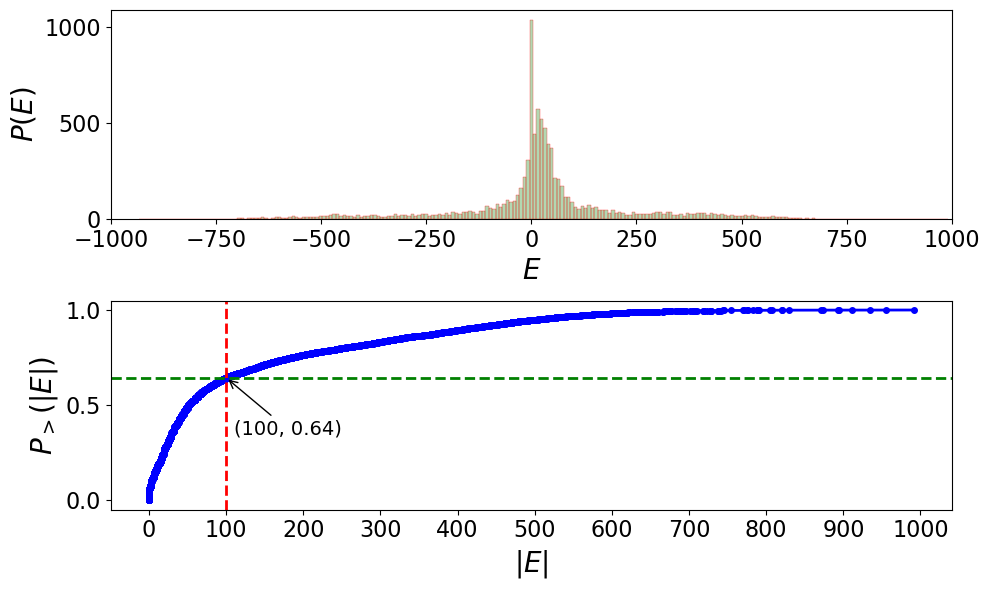}
        \caption{(Top) Probability to have a Stockfish $E$ (in centipawns). (Bottom) Cumulative distribution of the absolute value $|E|$. We observe here that positions with evaluation $|E|<100$ correspond to about $\approx 64\%$ of all positions (results computed for all positions for 100 games during the World Rapid chess championship 2023). }
	\label{fig:PE}
\end{figure}

The evaluation for all positions is distributed according to the result shown in Fig.~\ref{fig:PE}(top). The average is slighly positive as on average players with white win more often (the average is $\overline{E}=23$ and the median is $19$).
We also observe that a majority of positions during these games (about $64\%$) have an evaluation less than $|E|<100$ (Fig.~\ref{fig:PE}(bottom)).

\begin{figure}
   \includegraphics[width=0.4\textwidth]{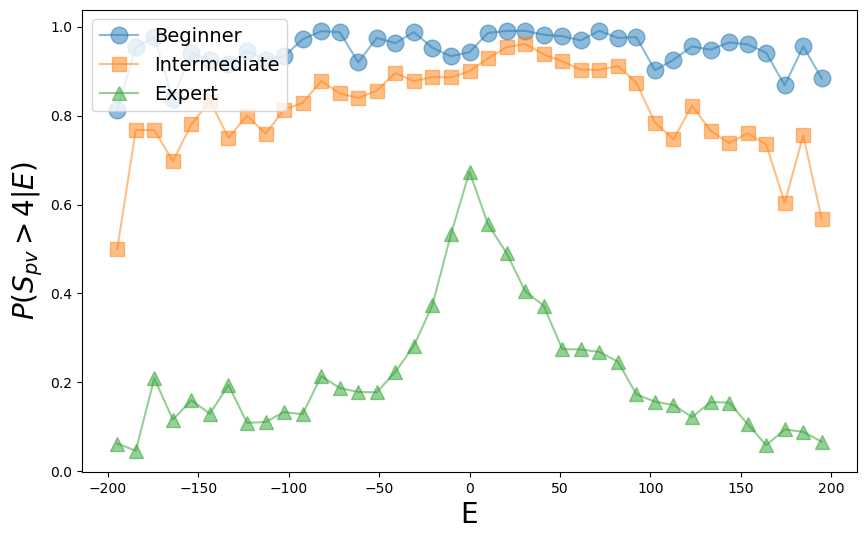}
        \caption{Probability to have a position with $S>4$ versus the engine evaluation $E$ for different levels. Computed for 100 games during the World Rapid chess championship 2023 (Samarkand). }
	\label{fig:PS4}
\end{figure}

An important question is then what is the entropy of the principal variation for a given value of $E$. We expect that for large $|E|$ the path to victory is made of moves relatively easy to find, while for smaller values, it might be more difficult. From the scatter plot $S$ versus $E$ (not shown), we compute the probability $P_{pv}(S>4|E)$ that the entropy of the principal variation is larger than 4 for a given value of $E$ (in practice, we bin the $E-$axis and for each bin count the fraction of position with an entropy $S>4$). The result is shown in Fig.~\ref{fig:PS4}.
This figure exhibits approximate symmetry around $E = 0$, suggesting that positions favoring white (positive $E$ or black (negative $E$) generally present similar levels of complexity for all player types. Different behaviors are observed across the player levels. Beginners consistently experience the highest entropy across all positions, as evidenced by the high $P(S_{pv}>4|E)$ values. This indicates that even straightforward positions for higher-level players might pose significant challenges for beginners, which suggests the necessity to have more straightforward guidance in AI evaluations. Intermediate players display a similar trend, albeit with slightly lower probabilities, reflecting their improved ability to manage complex positions compared to beginners. For both beginners and intermediates, $ P(S_{pv}>4|E)$ remains generally high over a wide range of $E$ values. The relatively flat curves in the approximate range $[-400, 400]$ indicate that high entropy, signifying complex or difficult sequences, is common across a broad range of positions for these players, even for large values of $|E|$. In contrast, the expert curve is significantly different. It displays lower probabilities of high entropy across most $E$ values, indicating that experts navigate
 clearly advantageous or disadvantageous positions with more ease, as these situations often present more straightforward plans. Notably, there is a sharp peak near $E = 0$ (typically around values $|E|<100$), suggesting that experts find positions with equal evaluations particularly challenging, likely due to the increased number of viable move options. We confirmed that this behavior remains consistent with small variations in $\Delta_0$ for experts (typically up to values around $30$). This figure highlights the significant differences in how complexity is perceived and managed by players of different skill levels. For experts, the main challenge lies in balanced positions, while beginners and intermediates struggle across a broad range of positions. This underscores the importance of tailored strategies and training to address the specific challenges faced by players at different stages of their chess development.

When analyzing the data by distinguishing different chess openings (see Methods for details), we observe that the results remain consistent across various opening types. Regardless of the specific opening being played, the trends in entropy and player performance are similar. For beginners and intermediate players, high entropy persists across a broad range of evaluation values, indicating consistent challenges in navigating complex positions. Experts, on the other hand, continue to show lower entropy levels overall, with a distinct peak near equal evaluations. This consistency across different openings suggests that the observed patterns are robust and not significantly influenced by the specific choice of opening, highlighting the general applicability of our findings across a wide range of chess positions.

\section{Conclusion}

In the 1950s, visionaries like Shannon and Turing speculated about the possibility of computers playing chess, laying the groundwork for what would become a revolutionary field of study (see the collection of papers in \cite{Levy:1988}). Over seventy years later, many of the questions they posed have been answered, largely due to remarkable advancements in artificial intelligence. The most significant breakthrough came with the advent of machine learning techniques, exemplified by AlphaZero, which marked a new era of chess engines that far surpass human capabilities. These engines not only redefine competitive play but also challenge long-held theories and principles in chess, providing fresh perspectives on openings and strategy \cite{Sadler:2019}. The confluence of these powerful engines and vast databases containing millions of games at all levels now lays the foundation for a rigorous scientific study of chess—an endeavor that, despite over a century of discussion, remains in its early stages.

In this work, we explored another critical dimension introduced by these advanced engines: the potential for AI-assisted learning and practice in chess. The concept of a computer evaluating a position, once theorized by Shannon and Turing, is now a reality. However, a significant gap exists between the numerical evaluation provided by engines and the practical realization of the corresponding moves, which requires players to execute a precise sequence of optimal choices. Our findings suggest that these engine evaluations often have limited practical value for human players, and should be adapted for interacting with less-skilled entities \cite{Hamade}. The recommended move sequences frequently exceed the cognitive limits of most players, particularly non-experts, underscoring a crucial disconnect: while AI excels in analysis, the information it generates is not always accessible or actionable for humans.

To bridge this gap, it is essential to rethink how engine evaluations are communicated. Making these evaluations more intuitive and tailored to different skill levels could transform them from mere assessments into actionable insights. This refinement would ensure that AI's assistance truly enhances, rather than overwhelms, the human player. By incorporating considerations of player skill into the design of AI tools, we can offer more detailed guidance and simplified recommendations, particularly for beginners and intermediates, thereby making the benefits of AI more universally accessible.\\




{\appendices
  \section*{Data}

There are numerous online resources available for accessing chess game data. For our analysis, we exclusively used open-access data, such as those available from \cite{data1, data2}. Typically, for a given opening, player, or event, we have access to a substantial number of games (ranging from several hundred to a few thousand) in Portable Game Notation (PGN) format. This format includes metadata about each game (such as date, location, and opponent) and the moves recorded in standard chess algebraic notation, where the chessboard columns are labeled \texttt{a} to \texttt{h} and the rows from 1 to 8. For each color (black and white), the six different pieces are denoted by P=pawn, R=rook, N=knight, B=bishop, Q=queen, and K=king, with lowercase letters representing the black pieces.

Our primary focus was on data \cite{data2} from the World Rapid and Blitz Championships held in Samarkand, Uzbekistan, in December 2023, featuring the world’s top players, where Magnus Carlsen emerged victorious in both formats.

Additionally, we analyzed various openings using data from
\cite{data1}. We concentrated on the most frequently played openings
during the 2024 Candidates Tournament, which took place in Toronto,
Canada, from April 3–22, 2024. The most common openings during this
tournament were, in decreasing order of occurrence: C42 (Petroff), C65
(Ruy Lopez, Berlin Defense), B30 (Sicilian Defense), C70 and C77 (Ruy
Lopez, other variations), and C50 (Giuoco Piano). The ECO codes for
these openings are given as referenced in \cite{ECO}.

\subsection*{Software}

We used python-chess (v1.9.4) which is a chess library for Python, with move generation, move validation, and support for common formats \cite{pythonchess}.

In particular, it implements an easy-to-use Stockfish class to integrates the Stockfish chess engine (version 16) with Python \cite{pypisf}. We run all our simulations with depth ranging from $20$ to $30$.

The Stockfish evaluation is given in centipawn. The centipawn is the
unit of measure used in chess as measure of the advantage. A centipawn
is equal to 1/100 of a pawn, and therefore 100 centipawns = 1
pawn. This can be compared to the commonly assigned value of various
pieces: pawns = 1, knights = 3, bishops = 3, rooks = 5, queens = 9.

}

\end{document}